\documentclass[lettersize,journal,twocolumn]{IEEEtran}
\usepackage{graphicx}
\graphicspath{{Figures/}}
\usepackage{amsfonts}
\usepackage{amssymb}
\usepackage{amsmath}
\usepackage{color}
\usepackage{wasysym}
\usepackage{hyperref}
\usepackage{bbold}
\usepackage{algpseudocode}
\usepackage{placeins}
\usepackage{tabularx}
\usepackage{makecell}
\usepackage{braket}
\usepackage{qcircuit}
\usepackage[utf8]{inputenc}   
\usepackage[T1]{fontenc}      
\usepackage{lmodern}          

\usepackage{mathtools}

\definecolor{lightblue}{RGB}{136,163,209}

\usepackage{algorithm}
\usepackage{algpseudocode}

\usepackage{authblk}

\usepackage{cite}

\begin{document}

\author[1]{Matteo Vandelli}
\affil[1]{Quantum Computing Solutions, Leonardo S.p.A., Via R. Pieragostini 80, Genova, 16151, Italy}
\author[1]{Francesco Ferrari}
\author[1,2]{Daniele Dragoni}
\affil[2]{Hypercomputing Continuum Unit, Leonardo S.p.A., Via R. Pieragostini 80, Genova, 16151, Italy}

\title{Constraint-preserving quantum algorithm for the multi-frequency antenna placement problem}

\maketitle

\begin{abstract}
Quantum algorithms for combinatorial optimization typically encode constraints as soft penalties within the objective function, which can reduce efficiency and scalability compared to state-of-the-art classical methods that instead exploit constraints to guide the search toward high-quality solutions.
Although solving this issue for an arbitrary problem is inherently a hard task, we address this challenge for a specific problem in the field of telecommunications, the \textit{multi-frequency antenna placement problem}, by introducing a constraint-preserving quantum adiabatic algorithm (QAA).
To this aim, we construct a quantum circuit that prepares an initial state comprising an equal superposition of all feasible solutions, and define a custom mixer that preserves both the one-hot encoding constraint for vertex coloring and the cardinality constraint on the number of antennas. This scheme can be extended to a broader range of applications characterized by similar constraints. We first benchmark the performance of this quantum algorithm against a basic version of QAA, demonstrating superior performance in terms of feasibility and success probability. We then apply this algorithm to large problem sizes with  hundreds of variables using a constraint-aware decomposition method based on the SPLIT framework. Our results indicate competitive performance against other large-scale classical approaches, such as branch-and-bound and simulated annealing. This work supports previous claims that constraint-aware algorithms are crucial for the practical and efficient application of quantum methods in industrial settings.
\end{abstract}

\section{Introduction}

Many combinatorial optimization problems arising in industrial applications~\cite{petropoulos2024or}, such as routing, scheduling, and resource allocation, are NP-hard and remain computationally challenging even for state-of-the-art classical algorithms like branch-and-bound~\cite{land1960bb,papadimitriou1982combinatorial,nemhauser1988integer} and simulated annealing~\cite{kirkpatrick1983sa}. These difficulties persist despite the enormous progress in high-performance computing (HPC).

Quantum computing approaches have been proposed as viable candidates for tackling these problems at scales currently intractable for classical digital computers, owing to their potential to significantly speed up solution space exploration~\cite{abbas2024review}. In this respect, quantum computers are particularly well-suited for solving quadratic unconstrained binary optimization (QUBO) problems~\cite{glover2022qubo}, a specific class of combinatorial optimization problems that can be directly mapped to Ising spin models~\cite{lucas2014ising} and natively encoded on quantum hardware.
Several industrially relevant combinatorial problems can be mapped to QUBO form by encoding constraints as soft penalties in the objective function. However, this formulation often introduces significant drawbacks: it can substantially increase the number of variables and generally requires careful, problem-specific penalty tuning to balance constraint enforcement against the original objective function. As a result, penalty terms often make the optimization landscape rougher, complicating the search of feasible, high-quality solutions~\cite{gabbassov2024lagrangian,shirai2025compressed}.

To address these challenges, various techniques have been proposed to extend the applicability of quantum algorithms to specific classes of constrained problems beyond QUBO~\cite{martonak2002tsp,hen2016driver,hen2016annealing}. In the context of the popular quantum approximate optimization algorithm (QAOA) \cite{farhi2014quantumapproximateoptimizationalgorithm, zhou2020qaoa, hadfield2019qaoa, blekos2024qaoa}, this can be achieved, for instance, by using Grover-like mixers~\cite{baertschi2020grover,xie2025performance}. General mixers satisfying linear constraints can be defined using the procedure outlined in Ref.~\cite{leipold2022constructing}, which nonetheless scales exponentially with the problem size in the case of general linear constraints. Some inequality constraints can be addressed using even more elaborate mixers based on quantum phase estimation~\cite{bucher2025ifqaoa}.

In this work, we study the \emph{multi-frequency antenna placement problem} (mAPP), an optimization problem with applications in Search-and-Rescue operations, and more broadly in the field of telecommunications. In mathematical terms, the mAPP translates into an unstructured quadratic program with binary variables and several equality constraints. We introduce a quantum adiabatic algorithm  specifically tailored to ensure feasibility under mAPP constraints, thanks to suitable mixer operators. We first benchmark the performance of this method on small mAPP instances by comparing its results to those of a QUBO-based quantum algorithm. Then, we leverage the SPLIT decomposition scheme~\cite{vandelli2025split} to define a hybrid quantum-classical algorithm that can tackle intermediate and large scale instances, with hundreds of variables. An extended comparison with various classical methods, such as branch-and-bound and simulated annealing, is carried out and highlights the effectiveness of constraint-aware approaches.

\section{Mathematical formulation of the problem}

\begin{figure}[t]
    \centering
\includegraphics[width=0.5\textwidth]{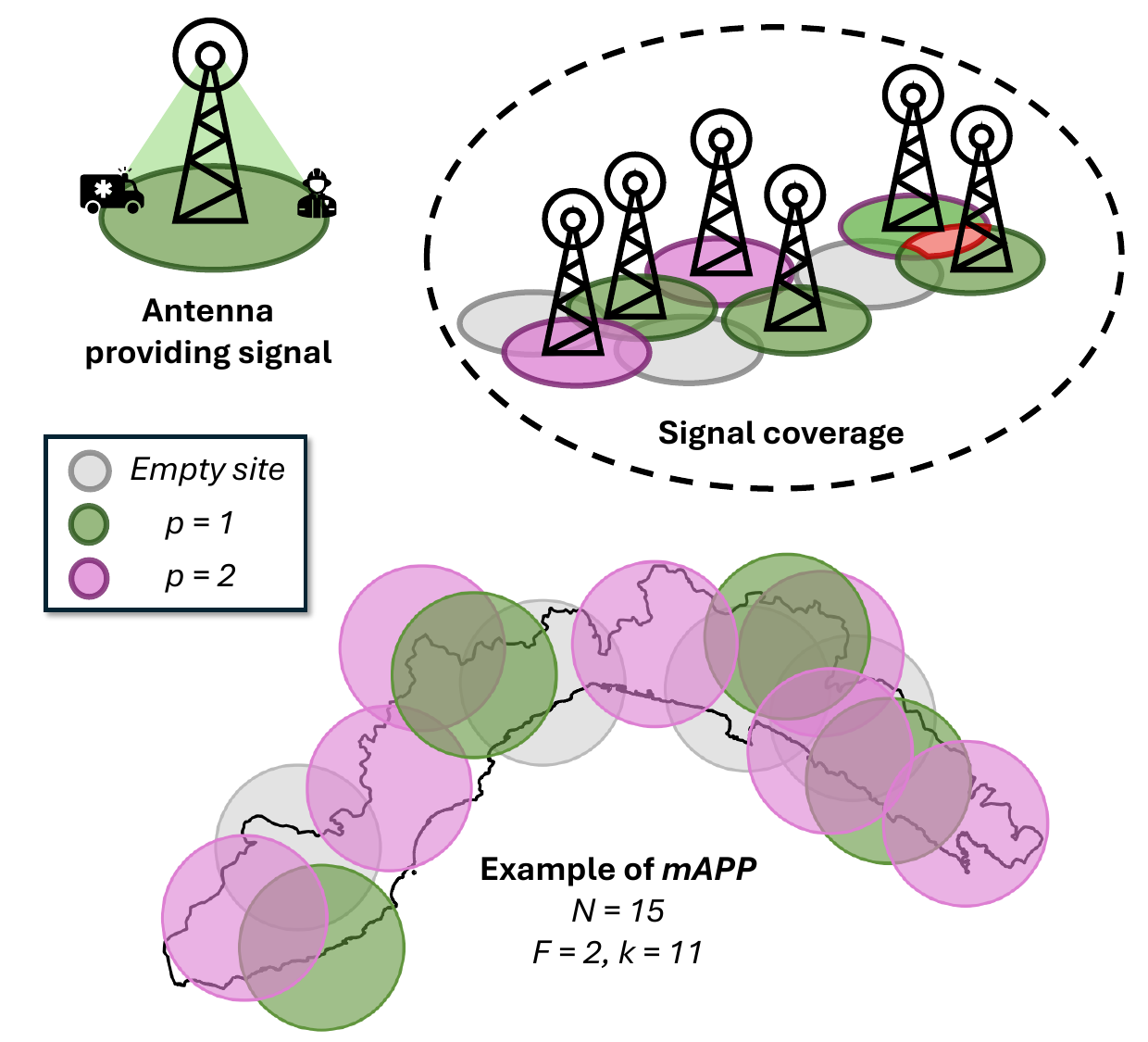}
    \caption{Top: schematic illustration of the multi-frequency antenna placement problem. Grey areas represent empty sites ($p=0$), while green/purple areas represent antennas operating at different frequencies ($p=1,2$). The red area indicates interfering signals. Bottom: solution of a mAPP instance on the Liguria region (Italy).}
    \label{fig:mapp}
\end{figure}

The multi-frequency antenna placement problem is motivated by the need to create and manage an \emph{ad-hoc} network of antennas on a territory hit by a natural disaster, providing signal coverage to the first responders on the ground (see Fig.\ref{fig:mapp}). In this scenario, a preliminary screening makes it possible to identify a set of potential locations at which to deploy a limited number of mobile antennas, that can be placed aboard vehicles. At that point, the available antennas have to be optimally placed on the territory in order to guarantee maximum coverage while avoiding interference. In order to achieve that, the operating frequency of the antennas can be chosen among different bands if there are overlaps between the antenna signals. The problem we formulate here aims to provide simultaneously the optimal placement of the antennas and the best frequency allocation in a given situation, providing a more realistic extension of the formulation given in Ref.~\cite{vandelli2024evaluating}

In mathematical terms, we consider the problem of placing $k$ antennas on $N$ candidate locations, in such a way to maximize coverage and minimize interference. We assume that each antenna can be chosen to operate at one of $F$ possible frequencies.
The problem has variables $x_{vp}$ where $v \in \{1, ..., N\}$ is the site index and $p \in \{0, 1, ..., F\}$ is the frequency index. The index $p=0$ corresponds to an empty site, thus has to be treated differently from the other $p$ indexes. We define $Q = N (F+1)$ to be the total number of binary variables of the problem in this one-hot encoding scheme. In the quantum methods adopted here, $Q$ qubits are used to encode these variables. The space of all the possible bitstrings is thus $S^{Q} = \{0, 1\}^{\otimes Q}$.

We define the cost function of the problem as 
\begin{align}
    C(x) = & \sum_{p,p'=1}^F \sum_{v<u} O^{(p, p')}_{vu} x_{vp} x_{up'} - \sum_{v = 1}^N \sum_{p=1}^F A_v x_{vp} \notag \\ &+ \alpha\sum_{v=1}^N\sum_{p=2}^{F} p \; x_{vp}
    \label{eq:cost_function}
\end{align}
where $A_v$ is the area covered by site $v$, $O^{(p,p')}_{vu}$ represents the interference due to the signal overlap between sites $v$ and $u$ when occupied by antennas operating at frequency $p$ and $p'$, respectively. In the last term, a small coefficient $\alpha$ is introduced to favor solutions that use fewer frequencies.
The binary quadratic program (QP) corresponding to this problem can be written as
\begin{align}
    \min_{x \in S^{Q}} &C(x)
    \label{eq:mAPP}\\
    &{\rm s.t.} \notag \\
    &\sum_{p=0}^F x_{vp} = 1, \;\;  \forall v
    \label{eq:one_hot_constraint}\\ 
    &\sum_{v=1}^{N}\sum_{p=1}^F x_{vp} = k
    \label{eq:number_constraint}
\end{align}
The first set of constraints indicates that each site can be either assigned an antenna with frequency $p\in\{1,\dots,F\}$ or be empty ($p=0$). The second constraint is the \emph{cardinality constraint} and enforces that the total number of antennas is $k$. The set of feasible solutions $S^Q_f$ satisfying these constraints contains exactly $|S^Q_f| = \binom{N}{k} F^{k}$ elements, as opposed to the total number of possible bitstrings $|S^Q| = 2^{N(F+1)}$.
This shows that the search space is decreased compared to an exhaustive search of all the possible bitstrings, but also that it can increase super-polynomially with the number of sites $N$ if $k$ grows with $N$. Note that, for given $N$ and $F$, the largest feasible subspace is achieved when $k=\lfloor \frac{F}{F+1}(N+1) \rfloor$.

\section{Quantum Methodology}

\subsection{Quantum adiabatic algorithm}

In the quantum approach, we solve the problem by encoding the variables $x_{vp}$ into a set of qubits $q_{v,p}$. The quantum adiabatic algorithm (QAA) solves optimization problems by smoothly evolving the quantum system of $Q$ qubits from an initial state $\ket{\psi_{\rm in}}$, which is the ground state of a \textit{mixer} Hamiltonian 
$H_{M}$, to the ground state of \textit{problem} Hamiltonian $H_{P}$, 
which encodes the optimal solution. This slow evolution is governed by the Hamiltonian
\begin{align}
   H(t) = (1 - s(t)) H_{M} + s(t) H_{P}. 
   \label{eq:qaa_continuous}
\end{align}
Here, $H_P$ can be, for instance, the Ising Hamiltonian obtained by quantization of the cost function of the problem \eqref{eq:cost_function}, namely $H_P = C\left(\frac{1 - Z}{2}\right)$, with $Z$ being the third Pauli matrix~\cite{lucas2014ising}. 

The scheduling function $s(t)$ satisfies $s(0) = 0$ and $s(T) = 1$, where $T$ is the total evolution time. In this work, we choose this function to be linear, namely $s(t) = t/T$.
The QAA algorithm exploits the fact that, after a sufficiently slow unitary evolution $U$ for a time $T$ under the action of Hamiltonian $H(t)$, the final state $\ket{\psi_{\rm out}} = U(T) \ket{\psi_{\rm in}}$ has a high probability of being the ground state of $H_P$~\cite{kato1950adiabatic,santoro2002qa,morita2008math,albash2018adiabatic}.
In the gate-based formulation of quantum computing, the continuous-time evolution of QAA is approximated by a discrete evolution, using the Trotter-Suzuki approximation, which yields a layered quantum circuit~\cite{farhi2000quantum, farhi2001quantum}. Indeed, discretizing the total runtime $T$ into $L$ equal steps of length $\tau = T/L$, the Trotterized time evolution at first order is given by
\begin{align}
U(T) \approx \prod_{l=1}^{L} 
\exp\!\Big(-i \, \tau \, \big(1-\tfrac{l}{L}\big) H_{M} \Big)\, \exp\!\Big(-i \, \tau \, \tfrac{l}{L} H_{P}\Big)
,
\end{align}
Different choices of the mixer and problem Hamiltonians are possible, depending on how constraints get incorporated into the QAA algorithm. This is discussed in the following two sections. 

\subsection{Basic QAA}\label{sec:qaa-basic}

The simplest version of the QAA (here denoted as QAA-BASIC) utilizes a local mixer, named the $X$-mixer, in the form $H_M = -\sum_{i=1}^Q X_i$, where $X_i$ is the NOT gate acting on qubit $i$. Its ground state is $\ket{\psi_{\rm in}} = \ket{+}^{\otimes Q}$, i.e. the uniform superposition of all $Q$-qubit bitstrings (written in the $Z$ basis). The algorithm in this form does not support hard constraints, so the quadratic program~\eqref{eq:mAPP} must be converted to the quadratic unconstrained binary optimization (QUBO) form~\cite{glover2022qubo}
\begin{align}\label{eq:qubo}
    \mathcal{Q}^{(\lambda)}(x) &= C(x) + \lambda \sum_{v=1}^N\left(\sum_{p=0}^F x_{vp} - 1\right)^2 \notag\\
    & + \lambda  \left(\sum_{v=1}^N \sum_{p=1}^F x_{vp} - k \right)^2,
\end{align}
In this formulation, the constraints of Eqs.~\eqref{eq:one_hot_constraint} and~\eqref{eq:number_constraint} have been replaced by soft penalty terms with coefficient $\lambda$. 
The corresponding problem Hamiltonian in the QAA becomes ${H_P = \mathcal{Q}^{(\lambda)}\left(\frac{1-Z}{2}\right)}$. It is convenient to normalize $H_P$ by dividing all its entries by the max norm.

\subsection{Constraint-preserving QAA}\label{sec:qaa-app}

For a constrained problem, the initial state $|\psi_{\rm in}\rangle$ need not be the uniform superposition of all possible bitstrings, but may be suitably devised 
to enforce feasibility~\cite{hadfield2019qaoa,wang2020xy}. In this case, the mixer $H_M$ should be selected so that $\ket{\psi_{\rm in}}$ is its ground state. 

The problem we address in this work is particularly amenable to the application of a constraint-preserving QAA. Indeed, we can construct a quantum algorithm that exactly preserves the constraints of the Eqs.~\eqref{eq:one_hot_constraint} and~\eqref{eq:number_constraint} by preparing an initial state $\ket{\psi_{\rm in}}$ that is an equal superposition of all feasible states, and a mixer operator $H_M$ that forces the system to remain in the feasible sector. This construction preserves both number and frequency constraints, so the problem Hamiltonian can simply be taken as ${H_P=C(\frac{1-Z}{2})}$ (rescaled by its max norm). The key ingredients of the constraint-preserving QAA, denoted in the following as QAA-APP, are:

\subsubsection{Initial state preparation}

\begin{figure}
    \centering

\[
\Qcircuit @C=1em @R=1.2em {
\lstick{ \begin{array}{c} q_{v,p} \\ \downarrow \end{array}  }  &  & & & \\
 \lstick{q_{1,0} \ }  & \multigate{1}{D_{N-k}^N} & \ctrlo{2} & \qw & \qw \\
 \lstick{q_{2,0} \ }  & \ghost{D_{N-k}^N}        & \qw       & \ctrlo{4} & \qw \\
 \lstick{q_{1,1} \ }  & \qw & \multigate{2}{W_R} & \qw & \qw \\
 \lstick{q_{1,2} \ }  & \qw & \ghost{W_F}        & \qw & \qw & \ \ \ R_1\\
 \lstick{q_{1,3} \ }  & \qw & \ghost{W_F}        & \qw & \qw \\
 \lstick{q_{2,1} \ }  & \qw & \qw & \multigate{2}{W_R} & \qw \\
 \lstick{q_{2,2} \ }  & \qw & \qw & \ghost{W_R}        & \qw & \ \ \ R_2\\
 \lstick{q_{2,3} \ }  & \qw & \qw & \ghost{W_R}        & \qw \\
 { \gategroup{4}{1}{6}{5}{1.32em}{--} 
  \gategroup{7}{1}{9}{5}{1.32em}{--} } }
\]

    \caption{Example of the state preparation circuit for the case of two sites with $F=3$, which creates the initial mAPP state of Eq.~\eqref{eq:feas_sup}. The block indicated with $D^N_{N-k}$ creates the Dicke state with $N-k$ Hamming weight on the first $N=2$ qubits. The $W_R$ blocks prepare the corresponding W states on the target qubits in the $R_1$ and $R_2$ registers. The empty circle indicates an inversely controlled gate.}
    \label{fig:prep_circ}
\end{figure}
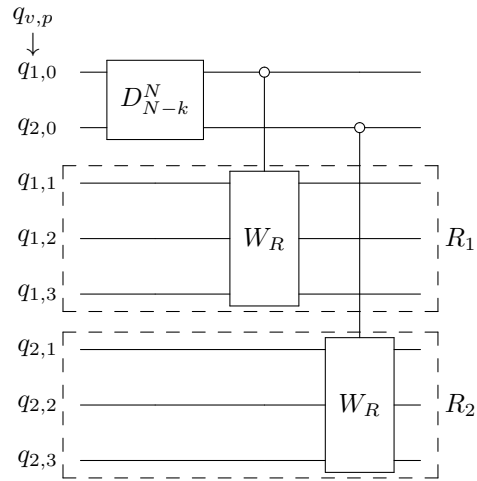

The aim of state preparation is to construct a uniform superposition of feasible solutions. Here we choose to take the equal superposition of all feasible bitstrings with $N$ sites, $k$ antennas and $F$ frequencies, i.e.
\begin{align}
    \ket{\psi_{\rm in}} = \frac{1}{\sqrt{|S_f^Q|}} \sum_{x \in S_f^Q} \ket{x}
    \label{eq:feas_sup}
\end{align}
To explain the circuit that prepares $\ket{\psi_{\rm in}}$, let us introduce a convenient ordering of the qubits, exemplified in Fig.~\ref{fig:prep_circ}. The first $N$ qubits are taken to represent the variables corresponding to the \emph{empty sites}, i.e. $q_{v, p=0}$ for $v \in \{1,\dots, N\}$. Then, $N$ groups of $F$ qubits are stacked consecutively in the register. Each of these groups, denoted by $R_v$, is formed by the qubits $\{q_{v, 1}, \dots, q_{v, F}\}$ for a certain site $v$. 

Since $k$ antennas need to be deployed in total, the feasible states are characterized by $N-k$ empty sites. Thus, the first step to prepare $\ket{\psi_{\rm in}}$ consists of creating a Dicke state~\cite{dicke1954coherence,nielsen2010quantum} with Hamming weight $N-k$ for the first $N$ qubits, i.e.
$\ket{D^{N}_{N-k}} \otimes |0\rangle ^{\otimes NF}$.
This can be achieved by using the circuit proposed in Ref.~\cite{baertschi2019dicke} with gate complexity $\mathcal{O}((N-k)N) \approx \mathcal{O}(N^2)$, which is represented by the $D^{N}_{N-k}$ block in Fig.~\ref{fig:prep_circ}. After that, we apply inversely-controlled $W$ gates, indicated as $\overline{CW}$, on the remaining qubits. This gate is defined as
\begin{align}
   \overline{CW}_{q, R} = \ket{0}\bra{0}_q \otimes W_R +  \ket{1}\bra{1}_q \otimes \mathbb{1}_R,
\end{align}
where $q$ is the control qubit and $W_R$ is the operator that creates a $W$-state on the qubits in the register $R$. The circuit that implements $W_R$ is given in Ref.~\cite{cruz2019W, diker2025W}. A series of $\overline{CW}$ gates is applied, each being inversely controlled on one of the first $N$ qubits $q_{v,p=0}$ and targeting the corresponding register $R_v$. The resulting state 
\begin{align}
    \ket{\psi_{\rm in}} = \left(\prod_{v=1}^N \overline{CW}_{v, R_v} \right)\ket{D^{N}_{N-k}} \otimes |0\rangle ^{\otimes NF}.
\end{align}
is exactly the equal superposition of all feasible states~\eqref{eq:feas_sup}. An example of the state preparation circuit is depicted in Fig.\ref{fig:prep_circ} for the case with $N=2$ and $F=3$. We use a version of $\overline{CW}$ that scales as $\mathcal{O}(F^2)$ in terms of gates, making the whole state preparation efficient with gate complexity $\mathcal{O}(N^2 + NF^2)$.

\subsubsection{Constraint-preserving mixer for mAPP}

The next important ingredient for the application of this algorithm is a constraint-preserving mixer. The mixer operator should enable exploration of the entire feasible subspace. This can be achieved by taking a two component mixer
\begin{align}
    H_{M} = H_{XY} + H_{\pm \pm}.
    \label{eq:mixing_op}
\end{align}
The first term is a regular XY mixer~\cite{hen2016annealing,hadfield2019qaoa} which, for each site $v$, acts on the $p>0$ frequency indices:
\begin{align}
    H_{XY} = 
    -\frac{\beta}{2}\sum_{v}\sum_{p=1}^F & \left(  X_{v,p} X_{v, (p\bmod F)+1} \right. \nonumber \\ 
    & \left. + Y_{v,p} Y_{v, (p\bmod F)+1} \right).
\end{align}
The second component has a $4$-qubit form similar to the permutation matrix previously introduced for vehicle routing problems~\cite{xie2024cvrp}, namely 
\begin{align}
    H_{\pm \pm} = -\beta\sum_{v<u} \sum_{p,p'=1}^F  \left(S^+_{v,p} S^+_{u,0} S^-_{v, 0} S^-_{u, p'} + {\rm h.c.}\right)
\end{align}
where $S^{\pm} =\frac{1}{2} (X \pm iY)$ are ladder spin operators. The operator $H_{XY}$ can increase or decrease the frequency index $p$ at each site $v$, keeping $p\neq 0$. The $4$-qubit term $H_{\pm \pm}$, instead, acts on pairs of sites and can swap an empty site ($p=0$) with a site operating at finite frequency. The combination of these two terms allows to explore all feasible bitstrings. Moreover, the initial state $\ket{\psi_{in}}$ of Eq.~ \eqref{eq:feas_sup} is the ground state of the Hamiltonian operator $H_M$ within the Hilbert subspace spanned by the feasible states. Hence, this construction is a consistent application of QAA.

From a gate-based perspective, the mixer $H_M$ presents two complications compared to the $X$-mixer. First, the operators in $H_M$ do not commute with each other. As a result, the finite-time evolution under $H_M$ is approximated using a Trotter–Suzuki decomposition (over $M$ steps). Following the scheme outlined in Appendix~\ref{app:trotter}, the feasibility of the solution is always preserved, regardless of the number of Trotter steps. Additionally, it requires the application of 4-qubit gates, which typically involves compilation in terms of 2-qubit gates. Each of the terms in the decomposition can be compiled using 6 CNOT gates and a few single qubit gates. The overall gate complexity of the mixer can be estimated as $\mathcal{O}(N^2F^2 M) = \mathcal{O}(Q^2 M)$, where $M$ is the number of Trotter steps. This increased circuit depth is partially balanced by the largely reduced search space $S^Q_f$, that allows for a reduced number of QAA layers $L$.  
We argue that these more complex state preparation and mixers result in circuit depths that will become practical only on fault-tolerant quantum hardware~\cite{eisert2025mindgap}.

\section{Computational methods}

\subsection{Classical methods}

To evaluate the performance of the above quantum algorithms, we compare the results against well-established classical methods. In Table~\ref{tab:methods}, we summarize the different methods employed in this work.

\begin{table}[t]
\renewcommand\arraystretch{1.5} 

\begin{tabularx}{\linewidth}{|>{\centering\arraybackslash}m{2.7cm}|X|}
\hline
Method & Description\\
\hline
\hline
\makecell{\vspace{0.1cm} \\ QAA-BASIC}   & Quantum adiabatic algorithm with local $X$-mixer and $H_P$ based on the QUBO form of mAPP~\eqref{eq:qubo}. \\
\hline
\makecell{\vspace{0.1cm} \\ QAA-APP}  & Quantum adiabatic algorithm with the constraint-preserving mixer~\eqref{eq:mixing_op} and $H_P$ based on the mAPP cost function~\eqref{eq:cost_function}.\\
\hline
\makecell{\vspace{0.1cm} \\ CPLEX} & \emph{State-of-the-art} branch-and-bound algorithm, as implemented in the IBM ILOG CPLEX solver~\cite{cplex2022v22}. \\
\hline
\makecell{\vspace{0.1cm} \\ SA} & Simulated annealing with local updates (single-variable flips), based on the QUBO formulation of mAPP~\eqref{eq:qubo}.\\
\hline
\makecell{\vspace{0.1cm} \\ CUSTOM-SA} & Simulated annealing with constraint-preserving updates, based on the original mAPP cost function~\eqref{eq:cost_function}.\\
\hline
\makecell{\vspace{0.2cm} \\ QAA-APP-SPLIT} & SPLIT method using QAA-APP as subproblem solver, with constraint-preserving SweepUdpate and update of the subproblem number constraints~\eqref{eq:local_constraint}.\\
\hline
\makecell{\vspace{0.cm} \\ CPLEX-APP-SPLIT} & \vspace{-0.65cm} SPLIT method using CPLEX as subproblem solver, with constraint-preserving SweepUdpate and update of the subproblem number constraints~\eqref{eq:local_constraint}. \\
\hline
\makecell{\vspace{0.cm} \\CPLEX-APP-SPLIT\\-PLAIN} & \vspace{-0.75cm} SPLIT method using CPLEX as subproblem solver as implemented in Ref.~\cite{vandelli2025split}, i.e. with single-variable SweepUdpate and static subproblem number constraints~\eqref{eq:local_constraint}.\\
\hline
\end{tabularx}
\caption{List of quantum, classical and hybrid quantum-classical methods used in this work, with their abbreviations.\label{tab:methods}}
\end{table}

The first classical approach is the branch-and-bound algorithm as implemented in the IBM ILOG CPLEX solver~\cite{cplex2022v22}, which guarantees optimality by systematically pruning the search space but has, in general, exponential worst-case complexity. In practice, when runtime becomes prohibitive, CPLEX can also be used as an approximate solver by imposing a time limit on its execution.

The second classical method considered in this work is simulated annealing~\cite{kirkpatrick1983sa}, a stochastic metaheuristic for local search that explores the energy landscape through probabilistic thermal fluctuations, often yielding high-quality solutions even for large-scale combinatorial problems. We evaluate the performance of two versions of simulated annealing. The first, referred to as SA, directly tackles the QUBO formulation of mAPP~\eqref{eq:mAPP}, in which constraints are implemented as soft-penalties. This variant relies on simple single-variable flips as update moves. On the other hand, the second version of simulated annealing, indicated as CUSTOM-SA, employs constraint-preserving updates that ensure the search always remains within the feasible subspace (once the system is initialized in a feasible state). The design of constraint-preserving moves essentially mimics the transitions between bitstrings allowed by the mixer Hamiltonian of Eq.~\eqref{eq:mixing_op}. Specifically, we introduce two kinds of moves. The first update rule allows to change the operating frequency of an active site: we extract a random site $v$ and identify the index $p$ at which $x_{v,p}=1$; if $p>0$, we propose the move $x_{v, p} \leftrightarrow x_{v, p'}$, with $p'>0$ a random frequency. The second update rule swaps the frequency indices of two sites: we randomly select two sites $v,u$ and identify $p_v,p_u$ giving $x_{v,p_v}=1$ and $x_{u,p_u}=1$; thus we propose the update $x_{v, p_v} \leftrightarrow x_{v, p_u}$ and $x_{u, p_u} \leftrightarrow x_{u, p_v}$. One possible effect of the latter move is to exchange $p=0$ with $p>0$ indices, effectively moving empty sites to different locations. 

This benchmarking framework allows us to compare the performance of QAA not only with a heuristic classical baseline, but also with a state-of-the-art exact solver, thereby assessing both the quality of the solutions and the scaling behavior with problem size.

\subsection{Problem decomposition (SPLIT)}

Emulating quantum algorithm on large-scale instances exceeding $30$-$40$ qubits remains computationally infeasible, even on modern HPC clusters. This effectively hinders the possibility of benchmarking these methods at industrially relevant scales. For this reason, in this work we combine the quantum methodology presented above with the SPLIT framework~\cite{vandelli2025split}, which enables the decomposition of large-scale problems into smaller subproblems that can be handled more easily by emulated quantum methods. This reduction also potentially enables execution on real quantum hardware, allowing algorithms to run with a number of variables that would otherwise exceed the capacity of current devices. 
Notably, the advantage of adopting the SPLIT pipeline is not limited to quantum methods, as it can also extend the applicability of classical branch-and-bound algorithms to problem sizes that would otherwise be intractable within reasonable computational time. While a thorough description of SPLIT is provided in Ref.~\cite{vandelli2025split}, here we outline the specifics and modifications applied in this work.

The first step of SPLIT consists of decomposing a mAPP instance into smaller mAPP subproblems. Let 
$\mathcal{C}$ denote the set of these subproblems. The variables $x_{v,p}$ of the original mAPP are assigned to the various subproblems $c \in \mathcal{C}$ via clustering. Specifically, we use \textit{spectral clustering} based on Euclidean distances to group sites, with each resulting cluster corresponding to a subproblem. Then, for each site $v$, all the variables $x_{v,p=0}, \dots, x_{v,p=F}$ are assigned to the corresponding subproblem $c$. This preserves the one-hot encoding structure of the mAPP variables in each subproblem.

Following subproblem decomposition, constraints are handled using the following strategy. One-hot constraints of Eq.~\eqref{eq:one_hot_constraint} can be imposed directly on each site $v$, thanks to the above clustering scheme that ensures $x_{v,p=0}, \dots, x_{v,p=F}$ belong to the same subproblem. On the other hand, the global number constraint \eqref{eq:number_constraint} is split among subproblems by imposing a local (i.e., subproblem dependent) constraint on the number of antennas
\begin{equation}\label{eq:local_constraint}
\sum_{v \in c}\sum_{p>0} x_{v, p} = k_c, \quad \forall c\in \mathcal{C}.
\end{equation}
Here $\{k_c\}_{c\in \mathcal{C}}$ are positive integer numbers that sum to the total number of antennas, i.e. ${\sum_{c} k_c = k}$. The initial choice of $k_c$ is taken to be proportional to the fraction of sites belonging to the subproblem $c$, as suggested in the original formulation of SPLIT~\cite{vandelli2025split}. However, in this work, we introduce an improved scheme which updates the values of $k_c$ at each iteration of the method. 

To this end, we develop a routine for the SweepUpdate step of SPLIT~\cite{vandelli2025split} that preserves the original constraints of mAPP. We use a greedy local search method based on the two types of moves previously introduced for CUSTOM-SA: the onsite move $x_{v, p} \leftrightarrow x_{v, p'}$ is applied to all sites $v$, while the two-sites move is applied to all the combinations of sites belonging to different clusters. Moves are accepted if they lead to a reduction in the cost function.
While both types of moves respect the mAPP constraints~\eqref{eq:one_hot_constraint}-\eqref{eq:number_constraint}, the two-sites move does not respect the local number constraints of Eq.~\eqref{eq:local_constraint}, as empty sites can be transferred between different subproblems. As a consequence, after the SweepUpdate step, the values of $k_c$ are updated to match the newly found configuration, by computing the number of antennas assigned to each subproblem $c$ in the current solution. This corresponds to a greedy update of the local number constraints~\eqref{eq:local_constraint} which reduces the dependence on the initial choice of $k_c$.

As summarized in Table~\ref{tab:methods}, we use SPLIT in combination with two different subproblem solvers, QAA-APP and CPLEX. In both cases, denoted by QAA-APP-SPLIT and CPLEX-APP-SPLIT (respectively), we adopt the constraint-preserving SweepUpdate routine and update the values of $k_c$ at each iteration. To perform a comparison with the original formulation of SPLIT~\cite{vandelli2025split}, we consider also the case of CPLEX-APP-SPLIT-PLAIN, in which $k_c$ are kept static and the SweepUpdate routine is based on single-variable flip updates.

\section{Results}

\subsection{Computational details}

We proceed by describing the main metrics used to evaluate the numerical results. We generate batches of $20$ mAPP instances for each problem size considered, with `problem size' referring to the total number of variables $N(F+1)$ (i.e., the number of qubits $Q$ in QAA methods). The different instances are constructed randomly generating candidate sites for antenna placement in the $20$ regions of Italy. We assume that interference takes place only between antennas operating at the same frequencies, i.e. $O^{(p,p')}_{vu} = O_{vu} \delta_{p,p'}$ in Eq.~\eqref{eq:cost_function}.

For what concerns QAA, we consider two metrics to assess the performance:
\begin{itemize}
    \item feasibility fraction, measuring the proportion of samples for which QAA is able to find a feasible solution to the problem
    \begin{align}
       p_{\rm feasible} = \frac{\# \; {\rm feasible \; counts} }{\# \; {\rm total \; counts}}
    \end{align}
    \item success probability, measuring the proportion of samples of QAA that correspond to one of the exact solutions of the problem
    \begin{align}
       p_{\rm success} = \frac{\# \; {\rm exact \; solution \; counts} }{\# \; {\rm total \; counts}}
    \end{align}
\end{itemize}

To compare the performance of different quantum and classical algorithms, we compute the
normalized difference 
    \begin{align}\label{eq:deltaalpha}
        \Delta \alpha = 1 - \frac{C(x_{\rm method})}{C(x_{\rm CPLEX})},
    \end{align}
    which quantifies the solution quality relative to the reference value provided by CPLEX. Lower values of $\Delta \alpha$ correspond to higher quality solutions. 
    Note that given the computational burden of large-scale instances, we set a time limit for CPLEX to $10$ minutes. This limit is never reached for instances with fewer than $400$ variables, which implies that, below this threshold, $x_{\rm CPLEX}$ coincides with the exact solution.

For methods based on CPLEX, the solution is obtained deterministically at the end of the calculation. In contrast, for stochastic methods, i.e.  QAA and simulated annealing, we examine all generated samples and select the best feasible solution. Specifically, the QAA solution corresponds to the best feasible bitstring obtained from $5000$ measurements. For simulated annealing approaches, the solution is taken as the best feasible result across multiple initializations: $1000$ for SA and $100$ for CUSTOM-SA.

As already mentioned, the reference against which we evaluate the performance of all methods is the CPLEX solver, through the ILOG CPLEX software package in Python~\cite{cplex2022v22}.
Regarding the quantum approaches, the QAA circuits are implemented using the \emph{Qiskit}  package from IBM~\cite{Qiskit}. The decomposition of the mixer in terms of common gates is obtained using the \texttt{transpile} method contained in this library. 
The simulated annealing methods exploit a custom implementation in Python using \emph{Numba} for just-in-time compilation~\cite{numbaref}.
Finally, the SPLIT framework uses the same Python implementation as in Ref.~\cite{vandelli2025split}, with the aforementioned improvements. The solution of the subproblems is performed in parallel using the \emph{mpi4py} package~\cite{mpi4pyref}. For the emulation of quantum circuits, we used a single node of our proprietary \emph{davinci-1} cluster equipped with AMD EPYC 7402 24-Core CPUs and NVIDIA A100 GPUs. The classical calculations were performed using only the node CPUs.

\begin{figure}[t]
    \centering
\includegraphics[width=0.5\textwidth]{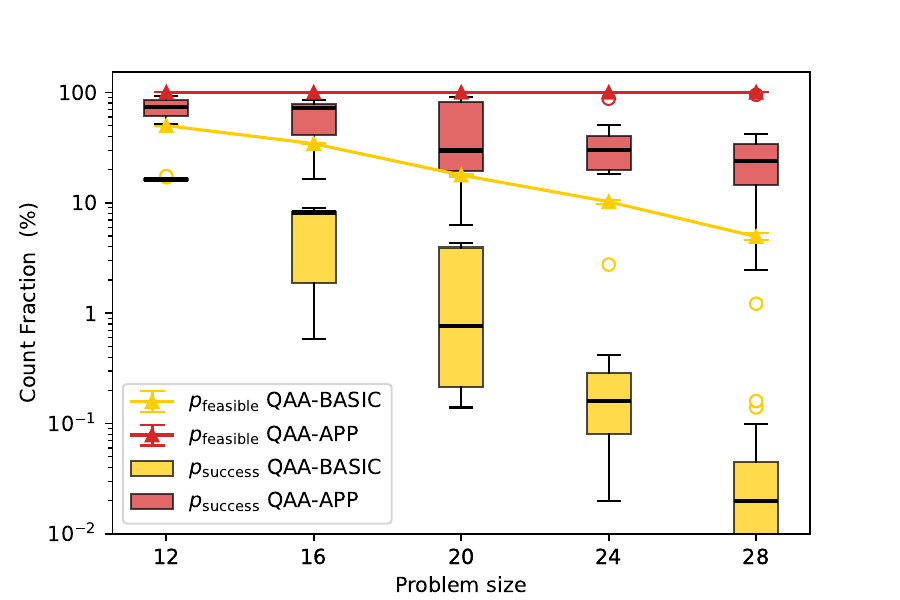}
    \caption{Average percentage of feasibility fraction ($p_{\rm feasible}$) of the various quantum methods tested on small instances (triangles and lines), superimposed to a boxplot showing the success probability ($p_{\rm success}$), in percentage. Each box shows the median and interquartile range, with dots indicating outliers. Instances with different problem sizes $N(F+1)$ are considered, with $F=3$ frequencies and $N=3,4,5,6,7$ sites. The number of antennas is set to  $k=\lfloor \frac{N}{2} \rfloor$.}
    \label{fig:feasibility}
\end{figure}

\subsection{Solution of small instances with QAA}

We begin by investigating small problem instances with up to $28$ variables, which can be solved directly using emulated QAA-BASIC and QAA-APP. For QAA-BASIC, we perform a Trotterization with $L=100$ steps. By analyzing a handful of instances, we determine effective values for the soft penalty term $\lambda$, chosen proportional to the maximum entry of the quadratic program coefficients, and for the total evolution time. The parameters found are then used for all simulations. For QAA-APP, we use $L=15$ Trotter steps and perform a similar analysis on a few instances to identify suitable values for the total evolution time and the mixer prefactor $\beta$. For what concerns the Trotterization of the mixer, we find that the value of $M$ does not impact significantly the performance of QAA-APP. Hence, we settle for $M=1$.

Our first objective is to examine the percentage of feasible solutions $p_{\rm feasible}$, shown in Fig.~\ref{fig:feasibility}. The plot highlights that the number of feasible bitstrings sampled by QAA-BASIC rapidly decreases as a function of $N$, while the QAA-APP retains by construction a 100\% feasibility across all  problem sizes. 
Specifically, for QAA-BASIC, we find a trend of $p_{\rm feasible}$ compatible with an exponential decrease as $N$ (and thus $Q$) increases; hence, for a fixed number of Trotter steps $L$, an exponential number of measurements would be required to sample a feasible solution at least once as the problem size increases. This finding extends previous work by demonstrating that, for highly constrained problem instances, in addition to an exponential decrease in the optimal-solution probability~\cite{vandelli2024evaluating,borle2021qaoa,willsch2022gpu,shaydulin2024evidence,montanez2025linear}, also the cumulative probability of feasible solutions can likewise become negligibly small.

After analyzing feasibility, we turn to the quality of the solutions obtained using QAA-BASIC and QAA-APP. A boxplot summarizing the success probabilities obtained in the various instances is shown in Fig.~\ref{fig:feasibility}. Our results show that $p_{\rm success}$ for QAA-BASIC decreases of several orders of magnitude in the range of problem sizes analyzed here, with a median trend compatible with an exponential function. This indicates that, in practice, identifying the exact solution with a limited number of measurements becomes extremely unlikely at large scales. On the other hand, the results for QAA-APP show a much slower decrease of $p_{\rm success}$, with the median value remaining above $20\%$ for the largest size considered here, compared to approximately $0.02\%$ for QAA-BASIC. Overall, this analysis of small instances demonstrates how a QAA evolution that preserves the problem constraints can lead to considerably superior performance. 

\subsection{Addressing larger problem sizes with SPLIT}

\begin{figure}
    \centering
    \includegraphics[width=0.5\textwidth]{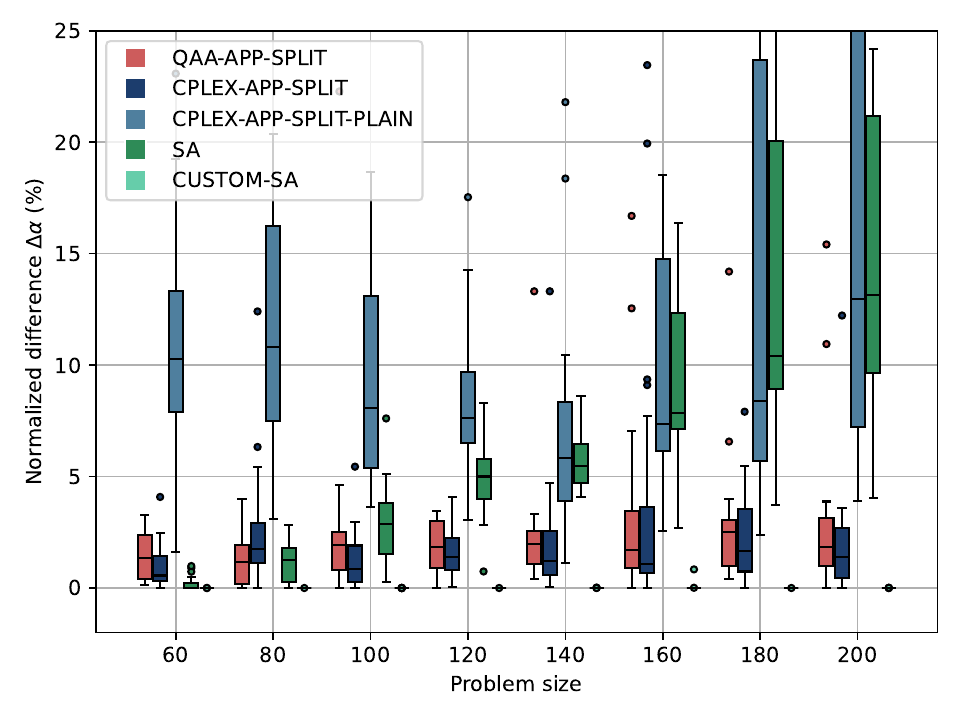}
    \caption{Boxplots of the normalized difference $\Delta \alpha$ [Eq.\eqref{eq:deltaalpha}] for different optimization methods (indicated in the legend) across intermediate problem sizes using hybrid quantum-classical and purely classical approaches. Each box shows the median and interquartile range, with dots indicating outliers.  Instances with different problem sizes $N(F+1)$ are considered, with $F=3$ frequencies and $N=15,20,25,30,35,40,45,50$ sites. The number of antennas is set to  $k=\lfloor \frac{F}{F+1}(N+1) \rfloor$.}
    \label{fig:quality_medium}
\end{figure}

In this section, we address intermediate instances of size up to $200$ variables. These instances cannot be directly solved with emulated QAA, but can be decomposed into smaller problems using SPLIT. This allows us to investigate the application of QAA-APP for larger problem sizes, which would otherwise be out of reach for emulation, and test them in combination with clustering approaches. Since QAA-BASIC can fail to provide feasible solutions already at $Q=28$, we consider here only QAA-APP in combination with SPLIT, which constitutes the hybrid method denoted as QAA-APP-SPLIT. We use a decomposition into $|\mathcal{C}|=6$ subproblems. We compare the results with different classical methods, namely SA and CUSTOM-SA, as well as CPLEX-APP-SPLIT and CPLEX-APP-SPLIT-PLAIN (see Table~\ref{tab:methods}). 

All of the above methods provide feasible solutions for all instances tested in this section. The results in Fig.~\ref{fig:quality_medium} show the quality of the solutions, as measured by the normalized difference $\Delta \alpha$ with respect to the exact result obtained by CPLEX.
Focusing first on simulated annealing methods, we observe that the performance of SA degrades roughly linearly as the number of variables increases, with median $\Delta \alpha$ exceeding $5\%$ at $140$ variables. On the other hand, the constraint-preserving CUSTOM-SA always finds the optimal solution in the whole range of variables considered here, showing the best performance among all methods. For what concerns methods based on the SPLIT decomposition, we observe that CPLEX-APP-SPLIT consistently outperforms its plain counter-part (CPLEX-APP-SPLIT-PLAIN), further indicating that constraint-aware methods for this kind of problems produce higher quality solutions. Furthermore, we note that using QAA-APP in place of CPLEX as subproblem solver yields roughly comparable performances in the whole range of problem size considered here (see results for CPLEX-APP-SPLIT and QAA-APP-SPLIT). This further confirms the reliability of suitably designed quantum algorithms as a viable alternative to state-of-the-art classical solvers.

\subsection{Scaling towards real-scale problem sizes}

To validate the various algorithms on problems of industrially relevant size, we test them on larger instances with up to $800$ variables. Note that in this regime, CPLEX also begins to struggle in both identifying and certifying the optimal solution within a reasonable computational time. Thus, to ensure consistency across experiments, we impose a time limit of $10$-minutes on the branch-and-bound solver. Therefore, as optimality is no more guaranteed, other methods may outperform the reference by finding better solutions than CPLEX, hence exhibiting a negative $\Delta\alpha$. It is worth noting that none of the other classical approaches exceeds the $10$-minutes time window for the instances considered here. 

\begin{figure}
    \centering
    \includegraphics[width=0.5\textwidth]{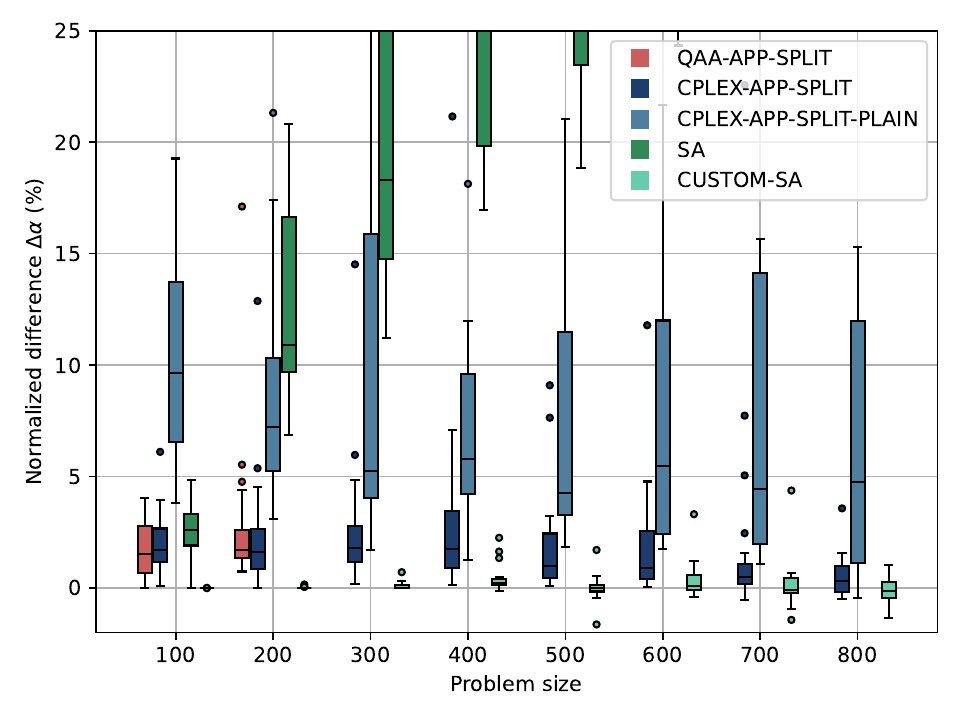}
    \caption{Boxplots of the normalized difference $\Delta \alpha$ for different classical optimization methods (indicated in the legend) across large problem sizes. Each box shows the median and interquartile range, with dots indicating outliers. The methods shown here always find a feasible solution. Instances with different problem sizes $N(F+1)$ are considered, with $F=4$ frequencies and $N=20,40,60,80,100,120,140,160$ sites. The number of antennas is set to  $k=\lfloor \frac{F}{F+1}(N+1) \rfloor$.}
    \label{fig:quality_large}
\end{figure}

The results for the normalized difference $\Delta\alpha$ are collected in Fig.~\ref{fig:quality_large}, for problem sizes ranging from $100$ to $800$ variables. We reiterate that, at this scale, a direct application of QAA is precluded by current emulation constraints.
Consequently, we show results of QAA-APP-SPLIT, decomposing each instance into $|\mathcal{C}|=9$ subproblems, as a comparison with the classical methods. On problem instances with $Q=100$ and $Q=200$, QAA-APP-SPLIT displays similar quality of the solutions as CPLEX-APP-SPLIT. However, as the problem size increases, one needs to proportionally increase the number of subproblems $|\mathcal{C}|$, to ensure that the size of the subproblems remains within the classical emulation limits. This, in turn, results in performance degradation due to the approximate nature of the decomposition scheme~\cite{vandelli2025split}.  Therefore, we assume that the performance of QAA-APP-SPLIT at larger problem sizes remains comparable to that of CPLEX-APP-SPLIT, and we proceed using only the latter method (still taking $|\mathcal{C}|=9$).

For what concerns classical methods, constraint-aware solvers (CPLEX-APP-SPLIT and CUSTOM-SA) perform largely better than the corresponding general-purpose solvers (CPLEX-APP-SPLIT-PLAIN and SA). Indeed, while the $\Delta \alpha$ of CUSTOM-SA always lies within $1\%$ difference from the CPLEX reference result, SA results consistently deteriorate with problem size. Analogously, CPLEX-APP-SPLIT-PLAIN exhibits a large variability in $\Delta\alpha$ on the instance pool (sometimes raising above $20\%$), signalling that a static distribution of the constraints $k_c$ is in general inadequate for this kind of problems. CPLEX-APP-SPLIT, instead, achieves better performance and even finds higher-quality solutions than CPLEX for a number of instances above $600$ variables. Overall, the trend of the data in Fig.~\ref{fig:quality_large} clearly indicates that constraint-aware approaches such as CPLEX-APP-SPLIT and CUSTOM-SA can outperform CPLEX at fixed runtime for large problem sizes.

\section{Conclusions}

In this work, we presented mAPP, a discrete optimization problem that combines antenna placement and frequency allocation in a single quadratic program. 
This problem exhibits prototypical challenges of real-world industrial cases, such as irregular cost-function structures and coefficients, and several constraints that must be satisfied to achieve a meaningful solution. We applied a pool of quantum, classical and hybrid algorithms to the solution of mAPP, assessing their performance. Some of these methods are designed to inherently handle problem constraints, while others require conversion to a QUBO-based formulation with soft penalties.

We performed a numerical analysis on three different ranges of problem sizes. We first focused on small instances, where we tested the performance of two variants of the quantum adiabatic algorithm, showing how a constraint-preserving algorithm for mAPP (QAA-APP) consistently outperforms the standard QAA implementation based on the QUBO formulation. 

We then moved to intermediate and large-scale instances, adopting the SPLIT metaheuristic~\cite{vandelli2025split} to extend the applicability of QAA-APP beyond the emulation limits on classical hardware. This yields the hybrid quantum-classical QAA-APP-SPLIT algorithm, which decomposes large problems into smaller subproblems, solved by QAA-APP. The numerical analysis on mAPP instances with hundreds of variables demonstrates that (i) the quality of the solutions obtained by QAA-APP-SPLIT mostly remains within $5\%$ from the optimal result, (ii) QAA-APP-SPLIT matches the performance of its fully classical counterpart, CPLEX-APP-SPLIT, in which the CPLEX branch-and-bound algorithm~\cite{cplex2022v22} is used as subproblem solver. This indicates the viability of quantum algorithms as an alternative to state-of-the-art classical solvers. Furthermore, comparison with other numerical methods highlights the importance of carefully handling constraints, especially for large-scale problems, as demonstrated also by the different performance of two simulated annealing variants, one QUBO-based (SA) \emph{vs} the other restricted to a local search within the feasible subspace (CUSTOM-SA).

Overall, our results challenge the widespread assumption in the quantum computing community that a straightforward deployment of QUBO-based quantum algorithms is sufficient to handle constraints in industrial applications. We emphasize that reformulating a quadratic program into QUBO form is often inadequate for problems with many constraints, as penalty tuning is instance-dependent and, even when done carefully, can still fail to deliver performance comparable to state-of-the-art classical methods. 

The present work on mAPP, as well as previous studies on vehicle routing~\cite{xie2024cvrp} and other problems~\cite{baertschi2020grover}, demonstrates that a problem-centric approach can achieve real-world utility, even before general-purpose approaches are developed to handle arbitrary constraints. Indeed, our QAA approach to the mAPP clearly indicates that the design of feasibility-preserving algorithms for specific problems can be formulated with relatively little effort compared to the case of generic constraints. As many combinatorial problems in the industrial sector are characterized by constraints similar to the ones discussed in this work, we argue that properly designed \emph{ad-hoc} state preparation and mixer circuits could be implemented for certain classes of common constraints, and incorporated into existing quantum algorithms, such as QAA or QAOA, to largely improve their effectiveness.

\section*{Acknowledgments}

This work was supported by the Next Generation EU program of the European Union through the Italian MUR National Recovery and Resilience Plan, Mission 4 Component 2 - Investment 1.4 - National Center for HPC, Big Data, and Quantum Computing (CN. 00000013 - SPOKE 10). We also gratefully acknowledge insightful discussions with the whole Quantum Computing Solutions team at Leonardo S.p.A., especially with Francesco Turro and Marco Maronese.

\section*{Disclaimer}
The authors declare no competing interest.

\appendices 
\section{Trotterization of the constraint-preserving mixer}\label{app:trotter}

The time evolution under the mixer of QAA-APP, Eq.~\eqref{eq:mixing_op}, is implemented by performing a Trotter decomposition over $M$ discrete steps, as follows
\begin{equation}
e^{-iH_{M} t} \approx \prod_{m=1}^M e^{-iH_{\pm \pm} \frac{t}{M}} e^{-iH_{XY} \frac{t}{M}} 
\end{equation}
Each term of the mixer is then further decomposed as
\begin{align}\label{eq:trotter_xy}
 &e^{-iH_{XY} \frac{t}{M}} \nonumber \\
 &\approx \prod_{v}\prod_{p=1}^F e^{i\frac{\beta t}{2M} \left( X_{v,p} X_{v, (p\bmod F)+1} + Y_{v,p} Y_{v, (p\bmod F)+1} \right) } \nonumber \\
 &= \prod_{v}\prod_{p=1}^F \left[e^{i\frac{\beta t}{2M}  X_{v,p} X_{v, (p\bmod F)+1} }\cdot  e^{i\frac{\beta t}{2M} Y_{v,p} Y_{v, (p\bmod F)+1}  }\right] 
\end{align}
and
\begin{align}\label{eq:trotter_pmpm}
 &e^{-iH_{\pm\pm} \frac{t}{M}} \nonumber \\
 &\approx \prod_{v<u}\prod_{p,p'=1}^F  e^{i \frac{\beta t}{M} \left(S^+_{v,p} S^+_{u,0} S^-_{v, 0} S^-_{u, p'} + {\rm h.c.}\right) } \nonumber \\
 &= \prod_{v<u}\prod_{p,p'=1}^F \left[
 e^{i \frac{\beta t}{8 M} X_{v,p} X_{u,0} X_{v, 0} X_{u, p'} }
 \cdot e^{i \frac{\beta t}{8 M} X_{v,p} Y_{u,0} X_{v, 0} Y_{u, p'} } \right. \nonumber \\
 & \phantom{\prod_{v<u}\prod_{p,p'=1}^F} \quad \cdot 
 e^{i \frac{\beta t}{8 M} Y_{v,p} X_{u,0} Y_{v, 0} X_{u, p'} }
 \cdot e^{-i \frac{\beta t}{8 M} X_{v,p} X_{u,0} Y_{v, 0} Y_{u, p'} } \nonumber \\
 & \phantom{\prod_{v<u}\prod_{p,p'=1}^F} \quad \cdot 
 e^{-i \frac{\beta t}{8 M} Y_{v,p} Y_{u,0} X_{v, 0} X_{u, p'} }
 \cdot e^{i \frac{\beta t}{8 M} X_{v,p} Y_{u,0} Y_{v, 0} X_{u, p'} } \nonumber \\
 & \phantom{\prod_{v<u}\prod_{p,p'=1}^F} \quad  \cdot \left.
 e^{i \frac{\beta t}{8 M} Y_{v,p} X_{u,0} X_{v, 0} Y_{u, p'} }
 \cdot e^{i \frac{\beta t}{8 M} Y_{v,p} Y_{u,0} Y_{v, 0} Y_{u, p'} }\right]
\end{align}
We observe that while a coarse Trotterization does not preserve the amplitudes of the bitstrings forming the superposition $\ket{\psi_{\rm in}}$, the order of the operators in the present scheme always guarantees that feasibility is preserved, even for $M=1$. Indeed the operators in the second rows of Eq.~\eqref{eq:trotter_xy} and Eq.~\eqref{eq:trotter_pmpm} are constraint-preserving. Their further decomposition (third rows of the above equations) is an exact equality, as all terms commute. Thus the final decomposition yields an operator that fully preserve mAPP constraints.

\bibliographystyle{unsrturl}
\bibliography{example}

\end{document}